# Neurocomputational Phenotypes in Female and Male Autistic Individuals

Michelle McCleod, Sean Borneman, Evie Malaia

Faculty Advisor: Evie Malaia

*Department of Communicative Disorders, The University of Alabama, Tuscaloosa, AL 35487*

**ABSTRACT:** Autism Spectrum Disorder (ASD) is characterized by an altered phenotype in social interaction and communication. Additionally, autism typically manifests differently in females as opposed to males – a phenomenon that has likely led to long-term problems in diagnostics of autism in females. These sex-based differences in communicative behavior may originate from differences in neurocomputational properties of brain organization. The present study looked to examine the relationship between one neurocomputational measure of brain organization, the local power-law exponent, in autistic vs. neurotypical, as well as male vs. female participants. To investigate the autistic phenotype in neural organization based on biological sex, we collected continuous resting-state EEG data for 19 autistic young adults (10 F), and 23 controls (14 F), using a 64-channel Net Station EEG acquisition system. The data was analyzed to quantify the 1/f power spectrum. Correlations between power-law exponent and behavioral measures were calculated in a between-group (female vs. male; autistic vs. neurotypical) design. On average, the power-law exponent was significantly greater in the male ASD group than in the female ASD group in fronto-central regions. The differences were more pronounced over the left hemisphere, suggesting neural organization differences in regions responsible for language complexity. These differences provide a potential explanation for behavioral variances in female vs. male autistic young adults.

**Introduction**

A hallmark of Autism Spectrum Disorder (ASD) is difficulties in social and communicative behavior; however, these difficulties are typically restricted to interactions between autistic and non-autistic individuals. Peer-to-peer communication between autistic individuals, by contrast, is just as efficient and enjoyable as that between neurotypical individuals [1]. The difficulty in communication between autistic and neurotypical individuals is thought to be derived from differences and incompatibility in neurotype. Human neurotypes may be quantified using neurocomputational measures of brain organization; one such measure is the representation of scale-free brain activity, termed the power-law exponent. In literature, a higher power-law exponent is associated with greater temporal variability in brain activity caused by an increase in the ratio of excitatory neurons to inhibitory neurons. Individuals on the autism spectrum[1] have previously been shown to have an increased excitation/inhibition ratio in their neural responses, compared to typically-developing individuals [3].

---

[1] Throughout the manuscript, the usage of both identity-first ("autistic") and person-first (individual with autism") language is intentional, recognizing that preferences may vary among individuals with Autism

However, autism frequently manifests differently between males and females. While symptoms more common in autistic males include problems in social-emotional reciprocity and repetitive behaviors, those more common in autistic females include internalizing behaviors and language mannerisms [4]. Additionally, the topics of restricted interests tend to differ across gender, with males more likely to show interest in particular objects or people, and females more likely to show interest in crafts or specific media. Autistic women are also much more likely to engage in camouflaging behaviors to mask their autism symptoms, blend in with those around them, and fulfill behavioral expectations [5–7]. The prevalence of female autistic camouflage is likely rooted in both biological differences, and, on a greater scale, cultural differences in treatment, acceptability, and expectations between males and females beginning in early development [8]. Because the diagnostic criteria for autism are biased towards observable atypical behaviors present in autistic males, many autistic females are under- and misdiagnosed, leading to mental health issues and problems in their personal and professional lives [7]. While the ratio of men to women diagnosed with autism is 4:1, research suggests the true ratio may be 3:4 [9].

Human brains generally show gender-based dimorphism in neural structure and function [10]. While female brains allow for more effective interhemispheric communication, facilitating connectivity between left and right hemispheres (i.e. left-lateralized language processing and right-lateralized social affect processing), male brain structure allows for optimal communication between neural regions within hemispheres [11]. Additionally, men and women tend to have significantly different patterns of neural activation during cognitive tasks, indicative of different cognitive processing strategies [12]. Endocrine regulation during development produces a significant effect on neural structure and function, with testosterone levels affecting gray matter volumes in limbic system structures in both sexes, and estrogen levels affecting regional brain activity in women [10,13]. Such differences, affecting both structure and function (possibly in a reciprocal manner) make it increasingly important to consider gender differences in analysis of both autistic and neurotypical populations.

This study aimed to objectively characterize gender- and diagnosis-based differences in resting state neural activity (i.e., awake with the eyes closed) in autistic vs. neurotypical males and females. Although the brain is always active in some way, the neural activity during the resting state is uniform enough to simulate a "baseline" for brain activity [14,15]. In this state, interregional communication between neural networks, thought to be the basis of cognition, can be observed using electrophysiological recordings such as electroencephalography (EEG). In EEG data analysis, differences in specific ranges of the power spectrum may reflect differences in dynamic network communication, which manifest as differences in behavior [3].

Raw EEG data is a combination of two signal components: one with oscillatory patterns and the other with arrhythmic 1/f patterns. The two distinct signal types are hypothesized to arise from distinct neural sources: oscillatory activity (task-related or resting-state), and background neural activity, which is related to cognitive processing indirectly by manifesting the multi-scale organization of neural networks and their connectivity. Because of this distinction, the oscillatory component tends to act as a confounding variable in studies investigating the 1/f pattern and vice versa [16]. To separate these two components and allow us to examine the 1/f component in isolation, we used a method called Irregular Resampling Auto-Spectral Analysis (IRASA).

---

Spectrum Disorder (ASD), informed by the understanding that language choice can be deeply personal and reflective of individual identity [2].

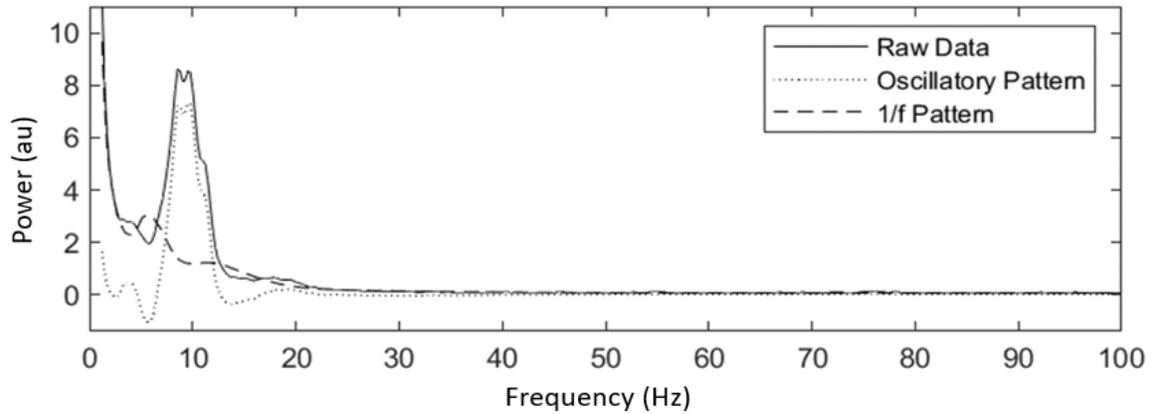

*Figure 1. EEG data distribution in raw recording*

IRASA separates the oscillatory and 1/f components by resampling the original signal using multiple non-integer pairwise factors. The calculation of the geometric mean of the auto-power spectra of every pair of resampled signals results in a new spectrum. This spectrum shows shifts in the power associated with the oscillatory component away from the original spectrum by certain amounts, depending on the resampling factor. Meanwhile, the power associated with the 1/f component maintains the same power-law distribution independent of the resampling factor. Taking the median of this spectrum results in the power spectra of the 1/f component. The oscillatory power spectra can be estimated by subtracting the 1/f power spectra from the mixed signal. The 1/f spectrum was transformed to a log-log scale and the least squares estimation was used to fit a linear function to the spectrum. The slope of this fitted line is the calculated power-law exponent [17].

Prior research has shown that autistic individuals differ from neurotypical ones in brain network communication, providing a potential explanation for differences in behavior and cognition [18,19]. Behaviorally, ASD also presents differently between males and females, likely due to the combination of genetic, developmental-biological, and cultural factors that lead to different trajectories in neural organization underlying behavior [20]. The present study aimed to use a neurocomputational measure, the power-law exponent, to objectively characterize gender-based dimorphism in EEG of autistic and neurotypical participants, assessing topological distribution of gender-driven variation in neural activity in terms of explanatory power for behavioral differences between groups.

**Methods**

**Participants**

To investigate the autistic phenotype in neural organization based on biological sex, we collected continuous resting-state EEG data for 19 autistic young adults (10 F, age 18-27, M = 20.7, SD = 1.9), and 23 controls (14 F, age 18-27, M = 21, SD = 1.8), using a 64-channel Net Station EEG acquisition system. Two groups of adults were recruited to participate in the study: a group of neurotypical adults (NT), and a group of ASD-diagnosed adults (ASD). The participants, ages 18-30, were recruited from the students at the University of Alabama and UA-ACTS program for high-functioning students on the autism spectrum.

**Data acquisition**

After participants provide informed, written consent approved by the University of Alabama Institutional Review Board, they were fitted with high-density 64-channel EGI Hydrocel Geodesic Sensor net caps. Then, 5 minutes of continuous EEG data was collected during resting state conditions of eyes-open (EO) using a Netstation EEG acquisition system. EEG data was sampled at 1000Hz/channel, band pass filtered at 0.1–60 Hz, and referenced to the vertex during recording. Electrode impedances were kept below 50KOhm, consistent with manufacturer's instructions. Raw data were exported to MATLAB as .edf (European Data Format) files for further analysis.

**Data Analysis**

Using IRASA, the power-law exponent per electrode per participant was calculated. Correlations were determined through JASP. Two-tailed Student's t-tests with a p-value of 0.10 were performed for each electrode using four between-group designs: ASD vs. neurotypical (NT), female vs. male autistic participants, autistic vs. neurotypical females, and autistic vs. neurotypical males. Additionally, to mitigate for multiple hypothesis testing for a large number of parameters (64 electrodes, in male vs. female and autistic vs. neurotypical populations), and since the use of parametric statistics for analysis of neural data does not take into account baseline distribution of EEG signal [3], additional exploratory feature analysis was conducted using machine learning algorithms. To construct the data matrix for machine learning analysis, the power of 1/f fit for the non-oscillatory component of the signal for each of the electrodes was organized into a matrix with 64 features and 42 instances (19 autistic + 23 neurotypical participants). Recursive Feature Selection (RFE) algorithm was used to identify salience of specific parameters (electrodes) for participant neurotype classification.

## Results

*ASD vs. NT*

RFE analysis identified electrodes 5, 23, and 39 as those with the highest salience for classification of autistic vs. neurotypical and male vs. female participants (see Figure 3).

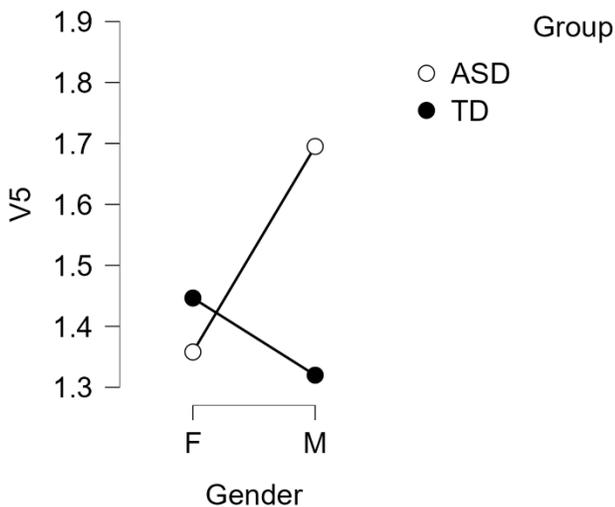

*Figure 2. Power-law exponent in EEG 1/f power spectrum over electrode 5 in male and female participants with autistic and non-autistic neurotypes (between-groups neurotype x gender ANOVA $F(1, 38) = 6.44$, $p<.015$, $\eta_p^2 = 0.145$).*

In contrast, EEG of ASD participants showed greater power-law exponent values (see Table 1) than those of the NT participants in Electrodes 1 and 17 (electrode 1, t-test $t(40)= 1.750$, $p<.088$, *Cohen's D = 0.543*; electrode 17, t-test $t(40)= 1.890$, $p<.066$, *Cohen's D = 0.586*). These electrodes detect electrical signals in lateral frontal regions of the brain and are placed symmetrically. No other comparisons have indicated statistical significance; $p>.1$.

| Electrode | Group | Mean | SD |
|---|---|---|---|
| 1 | ASD | 1.597 | 0.482 |
|   | NT | 1.379 | 0.323 |
| 17 | ASD | 1.556 | 0.369 |
|   | NT | 1.355 | 0.318 |

*Table 1. ASD vs. NT, power in the non-oscillatory 1/f signal component by electrode.*

### Female ASD vs. Male ASD

RFE analysis identified electrode 23 as one with the highest salience for classification of autistic males vs. autistic females. In contrast, in by-electrode statistical analysis, the values of power-law exponents were significantly greater in male ASD participants than female ASD participants in electrodes 2, 5, 10, 11, and 19 (Electrode 2, $t(17)= -2.021$, $p<.059$, *Cohen's D* $=-0.929$; Electrode 5, $t(17)=-2.107$, $p<.050$, Cohen's D$=-0.968$; Electrode 10, $t(17)= -1.837$ $p<.089$, *Cohen's* D$= -0.844$; electrode 11, $t(17)= -1.939$ $p<.069$, *Cohen's D* $= -0.891$; electrode 19, $t(17)= -1.989$ $p<.063$, *Cohen's D* $= -0.914$; see Table 2). Topologically, electrodes 2, 5, 10, and 11 detect electrical signals in the medial frontal regions of the brain bilaterally; electrode 19 is placed above the anterior portion of the left lobe. No other comparisons have indicated statistical significance; $p>.1$.

| Electrode | Group | Mean | SD |
|---|---|---|---|
| 2 | F | 1.252 | 0.239 |
|   | M | 1.564 | 0.403 |
| 5 | F | 1.358 | 0.290 |
|   | M | 1.695 | 0.393 |
| 10 | F | 1.335 | 0.251 |
|   | M | 1.616 | 0.392 |
| 11 | F | 1.234 | 0.175 |
|   | M | 1.409 | 0.213 |
| 19 | F | 1.162 | 0.196 |
|   | M | 1.332 | 0.177 |

*Table 2. Female ASD vs. Male ASD, power in the non-oscillatory 1/f signal component by electrode.*

### Female ASD vs. Female NT

Similarly to autistic male/female feature analysis above, RFE analysis identified electrode 23 as one with the highest salience for classification of autistic vs. neurotypical females. Statistical analysis for EEG of female participants also showed significantly greater power-law exponent values for ASD as compared to NT participants in electrode 23 (*t*(21)= 2.192 *p*<.040, Cohen's D= 0.937), placed above the left temporal lobe. However, power-law exponent values were significantly smaller in female ASD participants compared to female NT participants in electrodes 34 and 36, located on the midline over the parietal region (Electrode 34, *t*(21)= -1.750 *p*<.095, *Cohen's D* = -0.748; Electrode 36, *t*(21)= -1.822 *p*<.083, *Cohen's D* = -0.779). No other comparisons have indicated statistical significance; *p*>.1.

| Electrode | Group | Mean | SD |
|---|---|---|---|
| 34 | ASD | 1.192 | 0.200 |
|  | NT | 1.357 | 0.231 |
| 36 | ASD | 1.147 | 0.215 |
|  | NT | 1.304 | 0.192 |

*Table 3. Female ASD vs. Female NT, power in the non-oscillatory 1/f signal component by electrode.*

### *Male ASD vs. Male NT*

RFE analysis identified electrodes 5, 30, and 63 as those with the highest salience for classification of autistic vs. neurotypical males. Statistical by-electrode analysis for EEG for male participants showed significantly greater power-law exponent values for ASD as compared to NT participants in electrodes 1 (*t*(17)= 2.034, *p*<.058, *Cohen's D* = 0.935, ) and 17 (t-test *t*(17)= 2.486, *p*<.024x, *Cohen's D* = 1.142,), similar to the differences found in the ASD vs. NT field, as well as in electrodes 2 (*t*(17)= 1.983, *p*<.069, *Cohen's D* = 0.911) and 5 (*t*(17)= 2.640, *p*<.017, *Cohen's D* = 1.213), similar to the right hemispheric portion of the differences found in the Female ASD vs. Male ASD field. No other comparisons have indicated statistical significance; *p*>.1.

| Electrode | Group | Mean | SD |
|---|---|---|---|
| 1 | ASD | 1.649 | 0.446 |
|  | NT | 1.306 | 0.250 |
| 2 | ASD | 1.564 | 0.403 |
|  | NT | 1.268 | 0.205 |
| 5 | ASD | 1.695 | 0.393 |
|  | NT | 1.320 | 0.172 |
| 17 | ASD | 1.646 | 0.383 |
|  | NT | 1.297 | 0.184 |

*Table 4. Male ASD vs. Male NT, power in the non-oscillatory 1/f signal component by electrode.*

**Discussion**

In between-group analyses, the topography of salient electrodes tended toward the frontal regions of the brain, suggesting the role of executive- and limbic-system mechanisms in differentiating gender profiles in ASD. However, for the female-only comparison between autistic and neurotypical participants, it was the parietal and left temporal regions that yielded salient differences. The lack of measurable differences over the frontal lobe in autistic and neurotypical females correlates with behavioral data, which indicates that autistic and neurotypical women perform similarly in social cognition [8]. The differences over the temporal lobe might reflect sensory (auditory) differences between the two populations, as autists of both genders can be highly sensitive to a variety of sensory modalities. Additionally, language processing is often lateralized to the left Heschl's gyrus and Broca's area [21]. Because of the expected language-driven lateralization, the lack of symmetry among the salient electrodes over the left temporal region between ASD and neurotypical females is interesting and should be explored further [22].

In general, the data indicates that autistic males have the most pronounced differences from other groups studied. As a result, feature salience analysis for the entire data set is heavily influenced by data distribution in the autistic male group. This difference might be due to more articulated neurally-grounded differences in autistic males in general. On the other hand, autistic females, with higher interhemispheric connectivity [23] and different neuroendocrine developmental trajectory, as well as higher prevalence of camouflaging behaviors, differ from both autistic males and neurotypical females in neural organization. Focus on external symptoms of autism, more typical for males (repetitive behaviors, aggressions) and insufficient regard for less observable symptoms (sensory preferences, internalizing behavior, social attitudes/comfort) lead to failure to diagnose autistic women with accuracy, and maintenance of status quo - insufficient definition for clinical autistic profile in females

Parametric statistical analysis works under the assumption that the data follows a normal distribution [24]. EEG data, however, does not satisfy this assumption, as it follows a power-law distribution (see Figure 1). Thus, for data exploration, statistical analysis is not sufficient. We used machine learning technique – feature analysis – to identify electrodes that were salient for classification, following up with statistical analysis to ascertain directionality of the effect.

In the literature, a greater power-law exponent of the spectral profile of neural activity is associated with more temporally correlated coupling across neurons in brain circuits. On the other hand, a decreased power-law exponent is associated with increases in cognitive load [25]. Voytek & Knight (2015), based on intracranial recordings, hypothesized that autistic participants may show a flattened (decreased) averaged spectral slope compared to neurotypical ones. Our data revealed the opposite effect, suggesting that a more nuanced analysis of clinical populations is needed, and contrasting the contribution of dynamical neural activity that is localized (intracranial electrode patch) vs. global (full-scalp EEG) to the long-term behavioral profile in a clinical population.

In the step-down comparison between autistic and neurotypical males, the data indicated increased temporal correlation in the frontal region bilaterally in autistic, as compared to neurotypical males. For females-only analysis, the data indicated increased temporal correlation in the left temporal region and decreased temporal correlation in the medial parietal regions in autistic, as compared to neurotypical females. One potential explanation for the differences could be that autistic individuals, whose brain networks are highly sensitive to external stimuli [19], have developed sensory inhibition strategies that promote localized (within-region) brain activity over global (e.g. interhemispheric).

In terms of limitations, high between-participant variability in the study led to a lower signal-to-noise ratio, limiting inferential power of the analysis. This limitation might be mitigated by enrollment of a higher number of participants, especially females, in further studies, although the feasibility of it is problematic given the under-diagnosis of ASD in females.

**Conclusion**

The findings in our study shed light on the intricate neural underpinnings of Autism Spectrum Disorder (ASD), particularly in the context of gender differences. Between-group analyses revealed distinct topographical patterns, with frontal regions prominently implicated in differentiating gender profiles in ASD. Notably, autistic females exhibited significant differences in parietal and left temporal regions compared to neurotypical females, suggesting unique neural signatures potentially related to sensory processing and language perception. These findings challenge conventional assumptions about neural characteristics of ASD and underscore the importance of considering a broader range of symptoms beyond those traditionally associated with the disorder, especially in females who may exhibit camouflaging behaviors. In future studies, addressing the limitations of participant variability and under-diagnosis in females will be crucial for advancing our understanding of ASD and informing more accurate diagnostic and intervention strategies.